# ELEMENT ABUNDANCES AT HIGH REDSHIFTS: THE N/O RATIO IN A PRIMEVAL GALAXY

Max Pettini[1], Keith Lipman[2], and Richard W. Hunstead[3]

[1] Royal Greenwich Observatory, Madingley Road, Cambridge, CB3 0EZ, UK
(pettini@ast.cam.ac.uk)

[2] Institute of Astronomy, Madingley Road, Cambridge, CB3 0HA, UK
(kl@ast.cam.ac.uk)

[3] School of Physics, University of Sydney, NSW 2006, Australia
(rwh@astrop.physics.su.oz.au)

## ABSTRACT

The damped Lyman $\alpha$ systems seen in the spectra of high redshift QSOs offer the means to determine element abundances in galaxies observed while still at an early stage of evolution. Such measurements, which have only recently come within reach, complement and extend the data provided by studies of different stellar populations in our Galaxy and of extragalactic H II regions which have up to now formed the basis of galactic chemical evolution models. In this paper we demonstrate the potential of this new approach with high-resolution echelle observations of several elements in the $z_{\rm abs} = 2.27936$ absorption system in the bright $z_{\rm em} = 2.940$ QSO 2348−147. The absorbing galaxy appears to be chemically unevolved, with heavy element abundances only $\sim 1/100$ of solar; if it is the progenitor of a spiral galaxy like our own, it is unlikely to have collapsed to form a thin disk by $z \simeq 2.3$ (corresponding to a look-back time of $\sim 13$ Gyr for $H_0 = 50$ km s$^{-1}$ Mpc$^{-1}$ and $q_0 = 0.01$). Our data allow us to measure the nitrogen-to-oxygen ratio at a metallicity lower than those of the most metal-poor dwarf galaxies known. We find that, relative to the solar scale, N is more underabundant than O by at least a factor of $\sim 15$. This result is broadly in line with current ideas on the relative importance of primary and secondary production of N; future measurements in *several* damped Lyman $\alpha$ systems will permit more stringent tests of models for the evolution of the N/O ratio with time. Oxygen and other $\alpha-$elements are overabundant relative to Fe by no more than the factor of $\sim 3$ typical of metal-poor stars in the disk and halo of the Milky Way.

# 1. INTRODUCTION

The determination of the relative abundances of elements produced (and destroyed) through successive generations of stars is one of the major tools for following the evolution of galaxies from the earliest times to the present epoch. Until now data of relevance to the general question of galactic chemical evolution have been obtained mainly from two sources: studies of different stellar populations in the Milky Way and observations of bright H II regions in nearby galaxies. These two fertile avenues of work each suffer from their own limitations. Stellar abundances are normally deduced by comparing observed and theoretical spectra and therefore their accuracy depends to some extent on how successfully stellar atmospheres can be modelled. The stellar ages themselves are model dependent; this partly explains why it has proved difficult to establish a reliable 'clock' with which to measure the progressive enrichment of the Galaxy in heavy elements (Wheeler, Sneden, & Truran 1989). In any case, Galactic abundance studies are by definition parochial and, by themselves, tell us little about the chemical evolution of galaxies in general. Abundance determinations in extragalactic H II regions, do provide a broader view of this process in galaxies of different morphological types and masses. However, the range of elements accessible to ground-based emission-line spectroscopy is limited and the extent to which H II regions abundances are affected by local pollution can be a concern, particularly at low overall metallicities.

In this paper we draw attention to a different approach: the measurement of element abundances in the interstellar gas of high redshift galaxies as revealed by the absorption spectra of background QSOs. This technique has only recently begun to be explored in earnest but already shows great potential for complementing stellar and H II region abundance studies, particularly in the low abundance regime.

Traditionally, two main factors have limited the application of interstellar absorption line measurements to the study of chemical abundances in the Galaxy (Pettini 1985,

Jenkins 1987). First, under normal interstellar conditions significant fractions of most heavy elements are in solid, rather than gaseous, form and therefore do not give rise to discrete line absorption. Since the degree of depletion from the gas phase depends on the prevailing physical conditions and on the previous history of the dust grains, the necessary corrections to recover the overall (gas + dust) abundances can be both large and uncertain. Consequently, interstellar abundances, while providing an insight into the composition and processing of interstellar dust (e.g. Sofia, Cardelli, & Savage 1994) have been of limited relevance to Galactic abundance studies, although there are exceptions (e.g. Hawkins & Meyer 1989, Meyer et al. 1995). A second common complication arises from the limited dynamic range of interstellar absorption line measurements. In particular, for near-solar abundances many of the resonance lines of the most abundant astrophysical elements are too strong and saturated to be useful tracers of the corresponding column densities. For important species such as C and O, reliable abundance measurements are available only in a few directions towards stars sufficiently bright to show weak intercombination lines in their interstellar spectra (Cardelli et al. 1993).

Both difficulties can be alleviated when considering the class of QSO absorbers known as the damped Lyman $\alpha$ systems (Wolfe 1990). Several lines of evidence suggest that these absorption systems, which are characterized by high column densities of neutral hydrogen— $N(\mathrm{H}^0) \geq 2 \times 10^{20}$ cm$^{-2}$— may be the high-redshift progenitors of galaxies like our own. In particular, spiral galaxies dominate the absorption cross-section at such high columns of neutral gas in the local universe (Rao & Briggs 1993), and the mass density traced by the damped Lyman $\alpha$ lines at $z \simeq 3$ is comparable to that estimated to be in stars at the present epoch (Lanzetta, Wolfe, & Turnshek 1995). Taken together, these and other findings are at least consistent with the view that in the damped Lyman $\alpha$ systems we are seeing 'normal' galaxies at the time prior to the bulk of star formation, when most of their mass was in the interstellar medium. More direct evidence, as can be provided by imaging of the absorbers, is still patchy. While in some high-redshift cases large, luminous galaxies are indeed indicated (Briggs et al. 1989, Steidel & Hamilton 1992), at $z \leq 1$—where the imaging should be easier—the few damped systems identified to date appear to reside generally in underluminous galaxies (Steidel et al. 1994, 1995).

Although the question of which population(s) of galaxies give rise to damped Lyman $\alpha$ systems will probably not be settled until more extensive imaging surveys are carried out, it is already well established that these QSO absorbers trace gas at an early stage of chemical evolution. Several abundance studies (Pettini et al. 1994 and references therein) have shown that typical metallicities at $z = 2$ are 1/10 of solar, and that cases with $Z_{DLA} < 1/100 Z_{\odot}$ are not uncommon. Furthermore, dust depletions of refractory elements appear to be reduced significantly compared with the local interstellar medium and there are suggestions that, at the lowest values of $Z_{DLA}$, taking into account grain depletions to arrive at total element abundances may be a tractable problem (Pettini & Hunstead 1990; Fan & Tytler 1994; Lu et al. 1995). At the same time, in absorption systems with $Z_{DLA} \lesssim 1/100 Z_{\odot}$ the absorption lines of interest may be sufficiently weak (depending on the details of the velocity structure of the gas) to allow line saturation to be assessed adequately from high-resolution spectra, and reliable abundance estimates deduced.

The most metal-poor among the damped Lyman $\alpha$ systems then offer the unique opportunity to sample at first hand the mix of elements produced by the initial cycles of star formation in galaxies caught in their infancy. For $H_0 = 50$ km s$^{-1}$ Mpc$^{-1}$ and $q_0 = 0.01$ (this being the set of cosmological parameters which is least discordant with stellar ages) $z = 2$ corresponds to a look-back time of $\sim 13$ Gyr. Relative abundances measured in the interstellar media of these distant galaxies can therefore be compared directly with the record of such early stages in the chemical evolution of our own Galaxy preserved in the composition of old halo and disk stars. Whether in the outer regions of spirals or in irregular galaxies, H II regions with abundances less than 1/10 of solar are rare and none are known with $Z_{\rm H\ II} \lesssim 1/50 Z_\odot$, possibly because this is the minimum local enrichment produced by the same massive stars which ionize the gas and make the H II region visible (Kunth & Sargent 1986). The damped systems offer the means to extend the baseline of such abundance studies to lower metallicities by over one order of magnitude and be free from luminosity selection effects, since the high-redshift galaxies probed are recognized only because they fortuitously happen to lie in front of a background QSO as viewed from Earth.

Among the problems which we expect to be tackled are:

1. Are oxygen and the $\alpha-$elements overabundant relative to iron as found in Milky Way stars with [Fe/H] $< -1$? [We adopt here the usual notation for comparing measured abundances with solar values, $[{\rm Fe/H}] \equiv \log_{10}({\rm Fe/H}) - \log_{10}({\rm Fe/H})_\odot$]. Further, does the spread in [O/Fe] increase at the lowest metallicities, as tentatively suggested by recent observations of extremely metal-deficient stars (McWilliam et al. 1995)?

2. What is the behaviour of the N/O ratio at [O/H] $< -1$? Can its variation be understood in terms of different nucleosynthetic processes for the production of nitrogen (primary and secondary), or is the present scatter in N/O amongst dwarf galaxies due to self-pollution within the H II regions (Garnett 1990)? Answering this question has implications for extrapolating to the primordial abundance of helium (Pagel & Kazlauskas 1992).

3. What of the abundance of carbon in metal-poor systems? Measurements in H II regions require space observations which have only recently become possible (Garnett et al. 1995), whereas it is relatively straightforward to deduce [C/H] in QSO absorption systems provided the overall metallicity is low.

The full potential of abundance studies at high redshift will only be realized with efficient echelle spectrographs on large telescopes. However, some of these goals are already within reach, as we show in this paper where we report abundances for a wide range of elements in the damped Lyman $\alpha$ system at $z_{\rm abs} = 2.279$ in the spectrum of the QSO 2348−147 and are able to draw interesting conclusions on the N/O ratio at metallicities $\sim 1/100$ of solar.

## 2. OBSERVATIONS AND DATA REDUCTION

The bright ($V = 16.9$) high-redshift $z_{\rm em} = 2.940$ QSO 2348−147 was discovered by Monk et al. (1995) from a visual search of objective prism plates centred on bright galaxies. The 1950.0 coordinates are: Right Ascension = $23^{\rm h}\ 48^{\rm m}\ 55^{\rm s}.4$, Declination = $-14^\circ\ 44'\ 29''$. Follow-up intermediate dispersion spectroscopy confirmed that a strong absorption feature near 3987 Å in the discovery spectrum is indeed a damped Lyman $\alpha$ line at $z_{\rm abs} = 2.279$.

The data used in the present study were obtained during several observing runs, summarized in Table 1. The main set of observations consists of echelle spectra secured between 1991 and 1993 with the University College London Echelle Spectrograph at the coudé focus of the Anglo-Australian Telescope using the Image Photon Counting System as the detector. The instrumental set-up was similar to that described by Pettini et al. (1990); in particular, the 79 grooves mm$^{-1}$ echelle grating was chosen to provide sufficient inter-order separation for accurate subtraction of the background signal. Portions of the blue and ultraviolet spectrum of Q2348−147 between 4842 and 3728 Å (echelle orders 47–60) were recorded with three overlapping settings of the spectrograph at a resolving power $R \simeq 43\,000$, corresponding to a $FWHM \simeq 7$ km s$^{-1}$. The total wavelength range covered extends from $\approx 15$ Å longwards of the Lyman $\alpha$ emission line to $\approx 35$ Å longwards of the Lyman limit in the rest-frame of the QSO. There are small gaps between successive echelle orders—ranging from 25 to 3 Å— because the free spectral range of the 79 grooves mm$^{-1}$ grating exceeds the length of the detector at these wavelengths. Despite its relatively low DQE ($\approx 15\%$), the IPCS is the detector of choice for this work. The lack of any significant instrumental noise and the effective discrimination of cosmic-ray induced events make it possible for the observer to recover the signal-to-noise ratio appropriate to the photon counting statistics of object and sky only. This is essential at the very low count rates which are an inevitable consequence of observing relatively faint objects, such as QSOs, at such high spectral

TABLE 1

JOURNAL OF OBSERVATIONS

| Telescope (1) | Instrument (2) | Detector (3) | Dates (4) | Wavelength Range (Å) (5) | Exposure time (s) (6) | Resolution (Å) (7) | S/N [1] (8) |
|---|---|---|---|---|---|---|---|
| AAT | UCLES | IPCS | 10 – 13 Aug 1991<br>26 – 28 Aug 1992<br>19 – 22 Aug 1993 | 4002 – 4842<br>3921 – 4743<br>3728 – 4382 | 50210<br>45270<br>81000 | 0.10 | 4 – 10 |
| AAT | RGO Spectrograph | Thomson CCD | 9 Dec 1991<br>4 Jun 1992 | 4810 – 6060 | 4800 | 2.9 | 40 |
| AAT | RGO Spectrograph | Tektronix CCD | 30 Sep 1992 | 6810 – 8420 | 4000 | 2.9 | 35 |
| AAT | RGO Spectrograph | Thomson CCD | 11 – 13 Oct 1990 | 6310 – 6955 | 29900 | 1.4 | 80 |
| WHT | ISIS | IPCS | 29 Sep 1990 | 3790 – 4615 | 3600 | 2.0 | 12 |

[1] Measured signal-to-noise ratio near absorption lines in the $z_{\rm abs} = 2.279$ damped system.

resolution with 4-m class telescopes.

The echelle spectra were reduced, calibrated, co-added and normalized to the underlying QSO continuum as described by Pettini et al. (1990). Emission lines from a Th-Ar comparison lamp, extracted and added together in the same way as the QSO spectra, provided a measure of the resolution. The *FWHM* of the instrumental profile—sampled with $\sim 2.5$ wavelength bins—was found to be between 6.0 and 7.7 km s$^{-1}$ depending on the location of the lines on the detector (this is mostly due to distortions introduced by the magnetic focusing of the IPCS, rather than by the spectrograph optics). The signal-to-noise ratio of the final spectrum varies from order to order and along each order. The values we measured from the rms deviations of the data points from the fitted continuum near absorption lines of interest here are S/N $\simeq 4 - 10$. These values are lower than those generally obtainable with the UCLES+IPCS combination and the exposure times in Table 1; the cause of the reduced efficiency was later traced to a misalignment in the UCLES camera (now rectified) which resulted in additional vignetting.

The high-resolution echelle observations are augmented by intermediate dispersion CCD spectra obtained at the cassegrain focus of the AAT and covering three wavelength regions longwards of the Lyman $\alpha$ emission line (see Table 1). Two of these spectra were obtained primarily for the purpose of helping in the identification of metal-line systems in the Lyman $\alpha$ forest, but they are also of relevance to the work presented here because they include the expected locations of absorption lines in the damped system at $z_{\rm abs} = 2.279$. While the resolution of these data would normally be inadequate for abundance determinations, the velocity structure of the gas at $z_{\rm abs} = 2.279$ turns out to be sufficiently simple, and the abundances so low, that useful information can in fact be extracted even from spectra at 1–2 Å resolution (see below). The third portion of the red spectrum, from 6310 to 6955 Å, covers the expected positions of the Zn II and Cr II multiplets at high S/N and was obtained as part of the survey for these species by Pettini et al. (1994). Finally, the column density of H I has been measured from an intermediate dispersion blue spectrum obtained with the IPCS and the cassegrain spectrograph of the William Herschel Telescope on La Palma (last row of Table 1).

The complete set of spectra is presented in a separate paper (Mar, Hunstead, & Pettini 1995) which deals primarily with the properties of the Lyman $\alpha$ forest in Q2348−147; here we limit ourselves to considering the absorption lines in the $z_{\rm abs} = 2.279$ damped system.

## 3. THE $z_{\rm abs} = 2.27936$ DAMPED Lyman $\alpha$ SYSTEM IN Q2348−147

Our spectra cover absorption lines from a moderately rich array of astrophysically important elements, including H, C, N, O, Al, Si, S, Cr, Fe, Ni and Zn. Table 2 lists the observed transitions (or upper limits) of species which are the dominant ion stages of these elements in H I regions. Rest wavelengths and $f$-values are from Morton (1991), augmented where appropriate by recent revisions as compiled by Tripp, Lu, & Savage (1995). In column 4 of Table 2 we give the measured values of equivalent width *in the observed frame*, $W_{\rm obs}$, together with approximate errors which reflect the noise in the line but do not include the uncertainty in the continuum placement. For non-detections we list $3\sigma$ upper limits to $W_{\rm obs}$ calculated over the velocity range spanned by the lines detected. Vacuum heliocentric values of $z_{\rm abs}$ are given in column 6 with the appropriate number of significant figures depending on the resolution of the data (column 7); values which are uncertain because of noisy line profiles are flagged by a colon. From the best observed lines (C II $\lambda1334$, O I $\lambda1302$, Si II $\lambda1304$, Si II $\lambda1260$, and Si II $\lambda\lambda1190, 1193$) we deduce a mean absorption redshift for the damped system

$$z_{\rm abs} = 2.27936 \pm 0.00002 \ (1\sigma)\,.$$

Figure 1 shows the normalized profiles of selected absorption lines plotted on a common velocity scale centred at this redshift. At our resolution there appears to be one major velocity component (near $v = 0$ km s$^{-1}$ in Figure 1). It is possible that future observations of weaker transitions will show more than a single absorbing 'cloud' contributing to the absorption. However, we see no evidence in the present data for asymmetric profiles, nor for systematic velocity shifts between different ions, which are normally the symptoms of multi-component structure within the lines. (As can be seen from Table 2 below, individual line centroids differ by $\pm1.5$ km s$^{-1}$ from the mean redshift; such differences are within the errors expected from the accuracy of the wavelength scale and the S/N of the spectra).

TABLE 2
ABSORPTION LINE PARAMETERS OF NEUTRAL-PHASE SPECIES

| SPECIES | $\lambda_{vac}$ (Å) | LOG($\lambda$F) | $W_{obs}$[1] (mÅ) | $\lambda_{obs}$[2] (Å) | $z_{abs}$ | RESOLUTION (km s$^{-1}$) |
|---|---|---|---|---|---|---|
| (1) | (2) | (3) | (4) | (5) | (6) | (7) |
| $H^0$ | 1215.6701 | 2.704 | ... | 3986.1 | 2.2789 | 150 |
| $C^+$ | 1334.5323 | 2.232 | 565 ± 20 | 4376.44 | 2.27938 | 7.2 |
| $C^{+*}$ | 1335.7077 | 2.186 | < 82 [3] | ... | ... | 7.2 |
| $N^0$ | 1200.7098 | 1.725 | < 110 [3] | ... | ... | 7.6 |
| | 1200.2233 | 2.026 | < 110 [3] | ... | ... | 7.6 |
| | 1199.5496 | 2.202 | BLENDED | ... | ... | 7.6 |
| $O^0$ | 1355.5977 | −2.772 | < 60 [3] | ... | ... | 7.4 |
| | 1302.1685 | 1.804 | 480 ± 20 | 4270.25 | 2.27934 | 6.0 |
| $Al^+$ | 1670.7874 | 3.486 | BLENDED | ... | ... | 160 |
| $Si^+$ | 1808.0126 | 0.596 | < 250 [3] | ... | ... | 170 |
| | 1526.7066 | 2.248 | 500 ± 60 | 5006.1 | 2.2790 | 170 |
| | 1304.3702 | 2.089 | 335 ± 20 | 4277.49 | 2.27935 | 6.7 |
| | 1260.4221 | 3.148 | 530 ± 20 | 4133.40 | 2.27938 | 7.5 |
| | 1193.2897 | 2.831 | 410 ± 20 | 3913.21 | 2.27935 | 7.7 |
| | 1190.4158 | 2.530 | 385 ± 20 | 3903.78 | 2.27934 | 6.8 |
| $S^+$ | 1259.519 | 1.311 | 60 ± 15 | 4130.37: | 2.27932: | 6.8 |
| $Cr^+$ | 2066.161 | 2.027 | < 52 [3] | ... | ... | 60 |
| | 2062.234 | 2.206 | < 52 [3] | ... | ... | 60 |
| | 2056.254 | 2.334 | < 52 [3] | ... | ... | 60 |
| $Fe^+$ | 2382.765 | 2.855 | 670 ± 80 | 7813.6 | 2.2792 | 110 |
| | 2374.4612 | 1.889 | 290 ± 60 | 7786.6: | 2.2793: | 110 |
| | 2344.214 | 2.410 | 400 ± 70 | 7687.4 | 2.2793 | 110 |
| | 1608.4511 | 1.998 | 120 ± 45 | 5274.8: | 2.2794: | 170 |
| | 1144.9379 | 2.090 | 135 ± 25 | 3754.65: | 2.27935: | 7.0 |
| $Ni^+$ | 1454.842 | 1.938 | < 38 [3] | ... | ... | 6.7 |
| | 1317.217 | 2.284 | < 53 [3] | ... | ... | 6.7 |
| $Zn^+$ | 2062.664 | 2.723 | < 52 [3] | ... | ... | 60 |
| | 2026.136 | 2.996 | < 52 [3] | ... | ... | 60 |

[1] Equivalent widths in the *observed* frame of reference.
[2] Vacuum heliocentric wavelengths.
[3] $3\sigma$ detection limit for a $b = 10$ km s$^{-1}$ absorption feature.

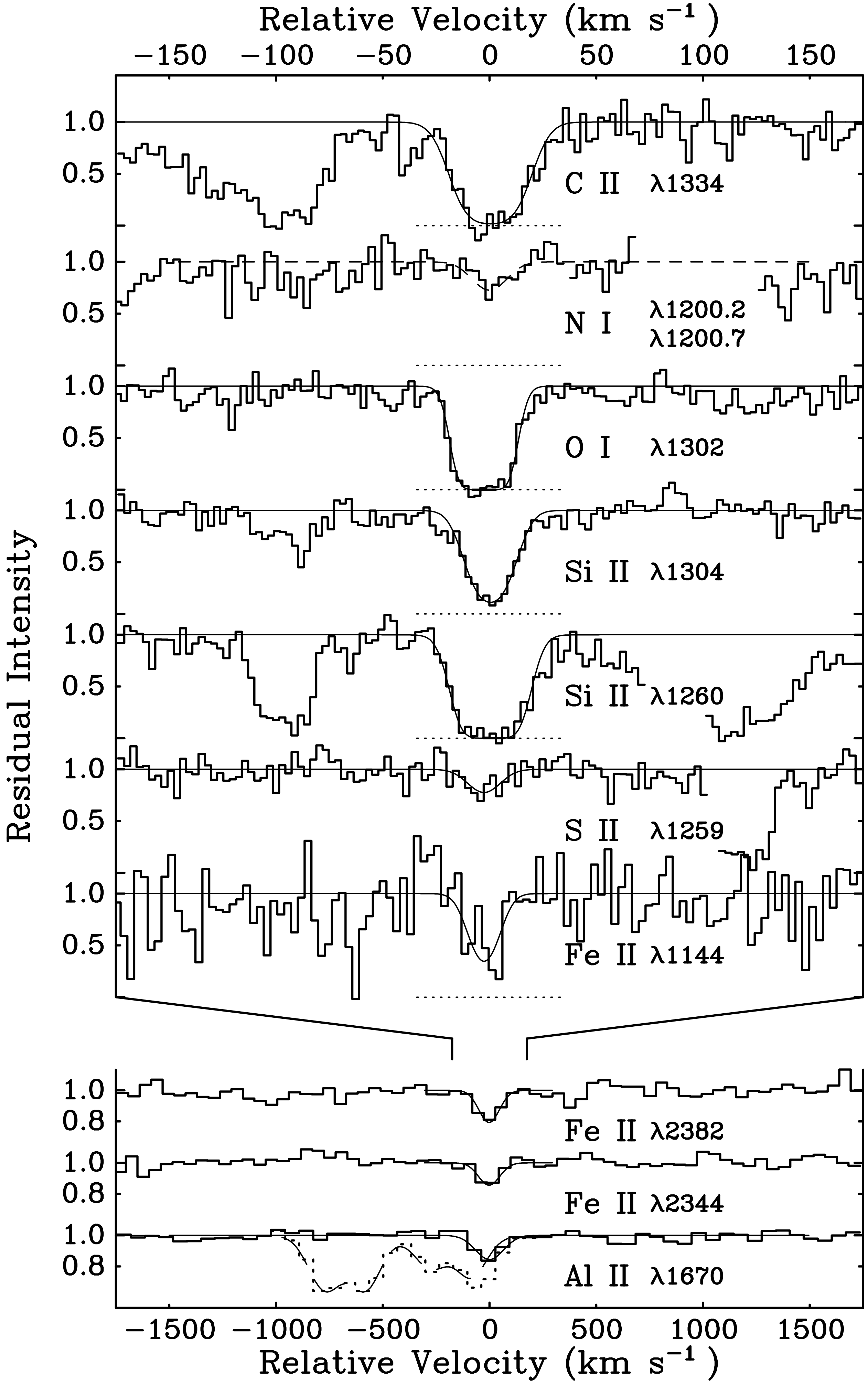

Relative Velocity (km s$^{-1}$)
−150 −100 −50 0 50 100 150
Residual Intensity
1.0
0.5
C II λ1334
N I λ1200.2 λ1200.7
O I λ1302
Si II λ1304
Si II λ1260
S II λ1259
Fe II λ1144
0.8
Fe II λ2382
Fe II λ2344
Al II λ1670
−1500 −1000 −500 0 500 1000 1500
Relative Velocity (km s$^{-1}$)

Inspection of Figure 1 also reveals that a second absorption feature is present near $v = -95$ km s$^{-1}$ in C II $\lambda 1334$ (where, however, it appears to be blended with other absorption within the Lyman $\alpha$ forest) and Si II $\lambda\lambda 1304, 1260$. The lack of any discernible component near this velocity in the well observed O I $\lambda 1302$ line indicates that in all likelihood this is H II gas.

We deduced ion column densities in the $v = 0$ km s$^{-1}$ component by comparing the observed absorption lines with theoretical profiles in the manner described by Pettini et al. (1990), using the interactive line fitting program *Xvoigt* (Mar & Bailey 1995) which is an improved version for Unix systems of the software used by Pettini et al.. Theoretical line profiles for a single Gaussian distribution of absorber velocities were computed and convolved with the appropriate instrumental broadening function; each model profile is uniquely defined by central velocity $v$, velocity dispersion parameter $b$, and column density $N$. For each absorption line, the three parameters were varied to explore the range of values of $N$ which are consistent with the measured profile and equivalent width. In cases where more than one transition from the ground state of the same ion was observed, the corresponding lines were analyzed simultaneously with the same set of parameters.

Figure 1: Normalized profiles of metal lines in the $z_{\rm abs} = 2.27936$ damped Lyman $\alpha$ system in the QSO 2348−147 plotted on a velocity scale centred on this redshift. Portions of the observed spectra have been reproduced as histograms, whereas the continuous thin lines are theoretical absorption profiles fitted to the data. Where useful, a dotted line shows the zero level. All the absorption features, apart from the bottom three, are from UCLES+IPCS observations at $\sim 7$ km s$^{-1}$ resolution; the last three are from intermediate dispersion CCD spectra and are shown an a wider velocity scale. The profile of N I is the composite of two transitions, as indicated; the long-dashed line shows the upper limit we deduce for the column density of neutral nitrogen. Al II $\lambda 1670$ (bottom of the figure) is blended with a C IV $\lambda\lambda 1548, 1550$ doublet at $z_{\rm abs} = 2.5320$, the observed and fitted profiles of which are indicated with broken lines.

Table 3

Profile Fit Parameters and Element Abundances in the $z_{\rm abs} = 2.27936$ Damped Ly$\alpha$ System

| Species | $v$ (km s$^{-1}$) | $b$ (km s$^{-1}$) | log $N$ (cm$^{-2}$) | log (X/H)$_{DLA}$ [1] | log (X/H)$_{\odot}$ [2] | [X/H]$_{DLA}$ |
|---|---|---|---|---|---|---|
| (1) | (2) | (3) | (4) | (5) | (6) | (7) |
| $C^+$ | $+0.5^{+1.5}_{-1.5}$ | $14.5^{+2.5}_{-10.5}$ | $14.38^{-0.11}_{+3.34}$ | $-6.19^{-0.11}_{+3.34}$ | $-3.44$ | $-2.75^{-0.11}_{+3.34}$ |
| $N^0$ | | | $< 13.47$ [3] | $< -7.10$ | $-3.95$ | $< -3.15$ |
| $O^0$ | $-2.5^{+0.5}_{-1.5}$ | $8.5^{+3.0}_{-3.5}$ | $15.37^{-0.58}_{+2.26}$ | $-5.20^{-0.58}_{+2.26}$ | $-3.07$ | $-2.13^{-0.58}_{+2.26}$ |
| $Al^+$ | $0.0^{+10}_{-10}$ | (10) [4] | $12.83^{-0.23}_{+0.24}$ | $-7.74^{-0.23}_{+0.24}$ | $-5.52$ | $-2.22^{-0.23}_{+0.24}$ |
| $Si^+$ | $+0.5^{+1.0}_{-1.0}$ | $10.0^{+1.0}_{-1.0}$ | $14.15^{-0.10}_{+0.13}$ | $-6.42^{-0.10}_{+0.13}$ | $-4.45$ | $-1.97^{-0.10}_{+0.13}$ |
| $S^+$ | $-2.5^{+2.0}_{-2.0}$ | $10.0^{+5.0}_{-3.0}$ | $13.93^{-0.16}_{+0.14}$ | $-6.64^{-0.16}_{+0.14}$ | $-4.73$ | $-1.91^{-0.16}_{+0.14}$ |
| $Cr^+$ | | | $< 12.49$ [5] | $< -8.08$ | $-6.32$ | $< -1.76$ |
| $Fe^+$ | $-2.5^{+3.0}_{-3.0}$ | $8.0^{+2.0}_{-2.0}$ | $13.73^{-0.10}_{+0.22}$ | $-6.84^{-0.10}_{+0.22}$ | $-4.49$ | $-2.35^{-0.10}_{+0.22}$ |
| $Ni^+$ | | | $< 12.86$ [5] | $< -7.71$ | $-5.75$ | $< -1.96$ |
| $Zn^+$ | | | $< 11.91$ [5] | $< -8.66$ | $-7.35$ | $< -1.31$ |

[1] Adopting $N(\mathrm{H}^0) = 3.7 \times 10^{20}$ cm$^{-2}$ (section 3.1).
[2] Solar system abundances from Anders & Grevesse (1989).
[3] See text.
[4] Adopted $b$ value; see text.
[5] 3 $\sigma$ detection limit for a $b = 10$ km s$^{-1}$ absorption feature.

The results of this profile fitting procedure are collected in columns 2, 3 and 4 of Table 3; sample fits are shown superposed on the observed metal lines in Figure 1. We now briefly discuss each ion in turn.

## 3.1. $H^0$

The damped profile of the Lyman $\alpha$ absorption line gives an adequately accurate measure of the neutral hydrogen column density: $N(\mathrm{H}^0)= (3.7 \pm 0.7) \times 10^{20}$ cm$^{-2}$ (Pettini et al. 1994). On the other hand, the precise redshift of the system is not well determined from the Lyman $\alpha$ line which, because of its strength, retains little information on the velocity structure of the H I gas. Consideration of the Lyman $\alpha$ profile alone yields a best estimate $z_{\mathrm{abs}}(\mathrm{Lyman}\ \alpha) = 2.2789$; this is the value reported by Pettini et al. (1994) and we have retained it in Table 2 here. However, as can be seen from Figure 2, the mean absorption redshift of the metal lines $z_{\mathrm{abs}} = 2.27936$ (+42 km s$^{-1}$ relative to $z_{\mathrm{abs}} = 2.2789$) gives an equally good fit within the uncertainties due to line blending in the Lyman $\alpha$ forest. (A velocity shift of 42 km s$^{-1}$ corresponds to approximately half a velocity bin in Figure 2).

Also reproduced in Figure 2 is the wavelength region near Si II $\lambda$1304 from the 7 km s$^{-1}$ resolution echelle spectrum; the portion of spectrum plotted is the same as that shown in the fourth panel of Figure 1. The comparison between the dramatically different velocity scales of the two absorption features highlights not only the need for high spectral resolution in metal abundance studies, but also the inherent difficulties in relating the column densities of heavy elements to that of neutral hydrogen. While the redshift agreement between metal lines and Lyman $\alpha$ is reassuring, we cannot exclude the possibility that there may be additional components which contribute to the broad Lyman $\alpha$ absorption but remain undetected in the metals. Such ambiguities can only be resolved by observations of higher order Lyman lines which, however, are beyond the range of the present data. The implications for our abundance determinations are discussed in section 4 below.

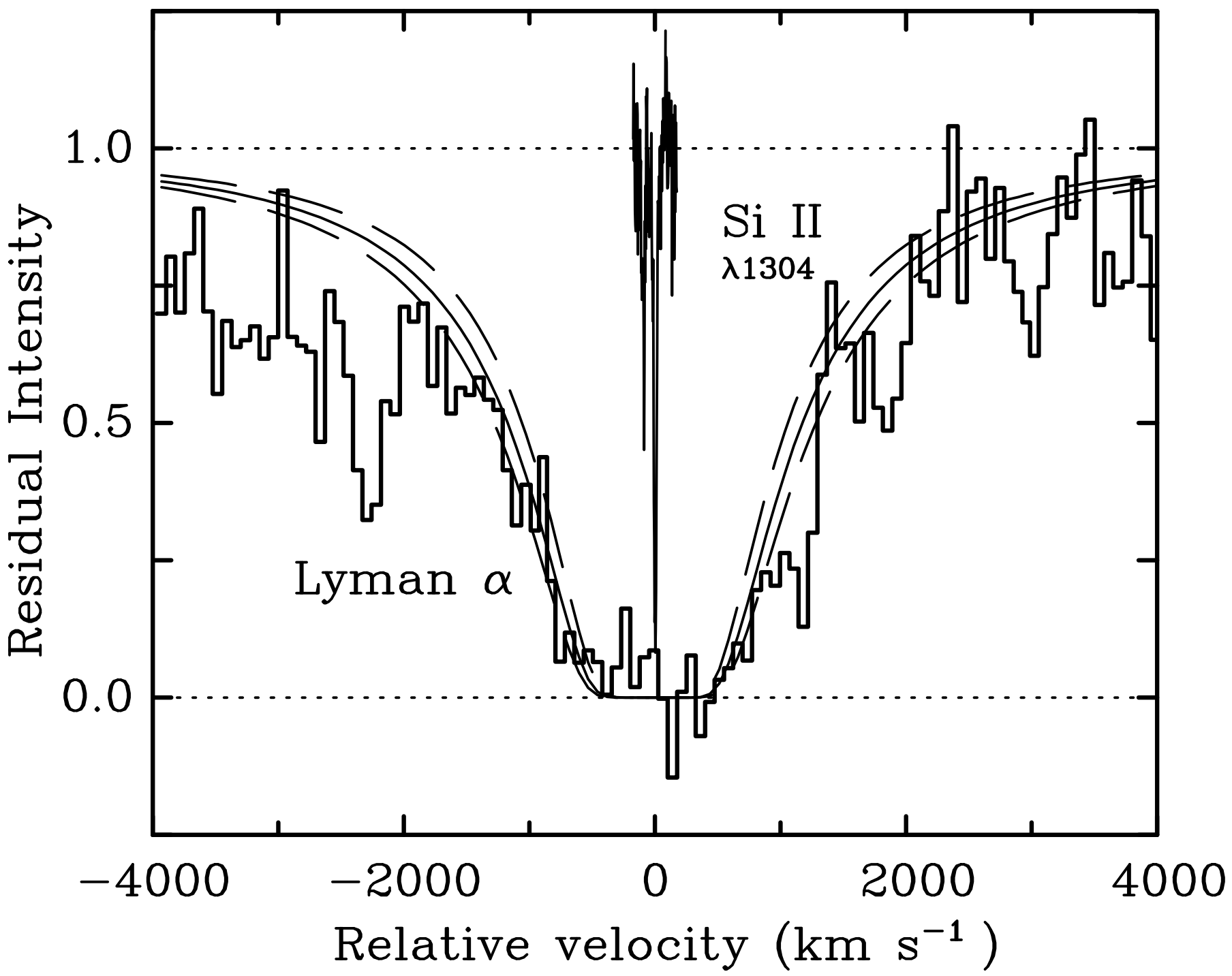

Figure 2: Profile fit to the damped Lyman $\alpha$ line in Q2348−147. The histogram shows the profile recorded at 2.0 Å resolution with ISIS+IPCS at the cassegrain focus of the WHT on La Palma; each bin is 1.0 Å (75 km $s^{-1}$) wide. The continuous lines show theoretical damped profiles centred at $z_{\rm abs}$ = 2.27936 (the mean redshift of the metal lines) for neutral hydrogen column densities $N(\mathrm{H}^0) = (3.7 \pm 0.7) \times 10^{20}$ $\mathrm{cm}^{-2}$. Also reproduced on the same velocity scale is the wavelength region covering the Si II $\lambda$1304 absorption line from the 7 km $s^{-1}$ resolution echelle spectrum obtained at the AAT.

### 3.2. $C^+$

Our spectra cover only one resonance line of singly ionized carbon, C II $\lambda$1334.5323. As can be seen from Figure 1, the line is saturated and the full range of $b - N$ solutions admitted by the line profile spans 3.5 orders of magnitude in column density (Table 3). Near the low $b$ end of this range the computed profiles approach the damping part of the curve of growth; while the corresponding values of $N(\mathrm{C}^+)$ are unreasonably high (in that they imply a carbon abundance greater than solar—see later), weak damping wings could in principle remain unnoticed in the present data. With better S/N spectra it will be possible to narrow the range of values of $N(\mathrm{C}^+)$ and thereby obtain more informative limits on the abundance of C.

C II* $\lambda 1335.7077$ is below our detection limit; we deduce a $3\sigma$ upper limit for the column density of $C^+$ in the $J = 3/2$ fine-structure level $N(C^{+*}) \leq 1.4 \times 10^{13}$ cm$^{-2}$.

### 3.3. $N^0$

We do *not* detect convincing N I absorption. In Figure 3 we have reproduced portions of the echelle spectrum encompassing the three lines of multiplet 1, N I $\lambda\lambda 1199.5496, 1200.2233, 1200.7098$. A feature with $W_{\rm obs} = 160 \pm 30$ mÅ is seen at $\lambda_{\rm obs} = 3933.68$, near the expected position of the strongest line $\lambda 1199.5496$, but the wavelength agreement is poor (the feature is blueshifted by 6 km s$^{-1}$ relative to the mean system redshift). Given that the centroids of the other metal lines agree with each other to within $\pm 1.5$ km s$^{-1}$, we consider it more likely that this is in fact Galactic Ca II $\lambda 3933.663$ at $v_{\rm H} = +1$ km s$^{-1}$. $W_{\rm obs} = 160 \pm 30$ mÅ is indeed typical of the values commonly measured for this line towards extragalactic sources (Bowen 1991).

The most sensitive limit to the column density of $N^0$ is obtained by adding together on a common velocity scale the wavelength regions including the expected locations of N I $\lambda\lambda 1200.2233, 1200.7098$. The resulting spectrum, reproduced in the second panel of Figure 1, shows a a $2.5\sigma$ feature at approximately the correct wavelength, the reality of which can only be confirmed with better S/N ratio data. We deduce a $3\sigma$ upper limit $N(N^0) \leq 3 \times 10^{13}$ cm$^{-2}$; the theoretical profiles shown in Figures 1 and 3 correspond to this column density for $b = 10$ km s$^{-1}$, the average velocity dispersion parameter of well-observed metal lines (see Table 3). As discussed below (section 4), the weakness of the N I absorption in this damped Lyman $\alpha$ system implies that nitrogen is underabundant by more than three orders of magnitude compared with the Sun.

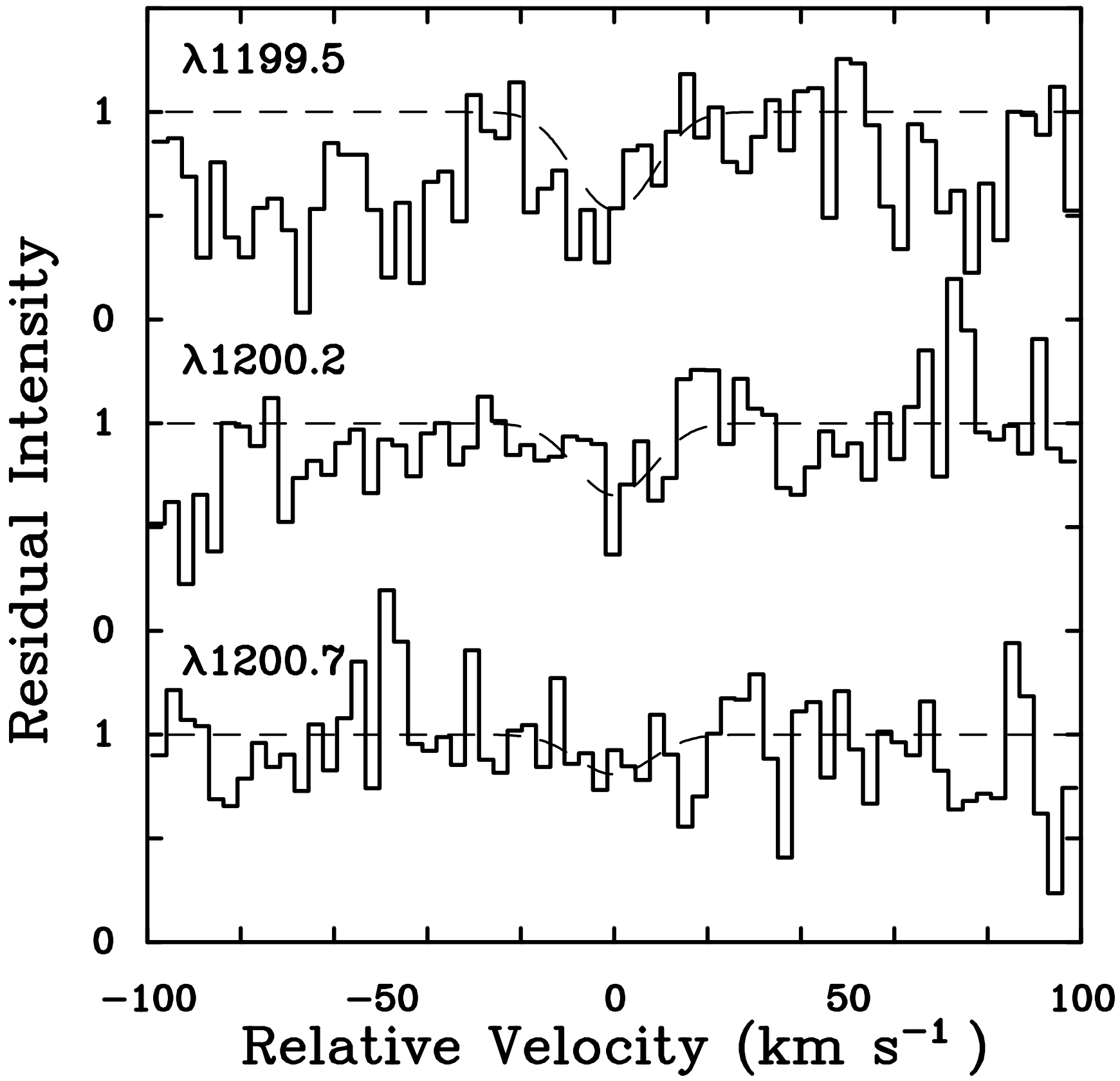

Figure 3: Portions of the echelle spectrum of Q2348−147 encompassing the expected positions of the three N I lines in multiplet 1 at $z_{\rm abs} = 2.27936$. The dashed lines show theoretical absorption profiles corresponding to the $3\sigma$ upper limit $N({\rm N}^0) = 3 \times 10^{13}$ cm$^{-2}$ and $b = 10$ km s$^{-1}$. The absorption feature blueshifted by a few km s$^{-1}$ relative to N I $\lambda1199.5496$ (top plot) is probably Galactic Ca II $\lambda3933.663$. The weakness of the N I absorption implies an underabundance by more than three orders of magnitude relative to the Sun.

### 3.4. $O^0$

Oxygen suffers from the same difficulties encountered for $C^+$—only one saturated resonance line is observed in our data. The solutions compatible with the absorption profile span the whole flat part of the curve of growth, corresponding to three orders of magnitude in abundance. We can only place a rather uninformative upper limit on the intrinsically weak ($f = 1.248 \times 10^{-6}$) O I $\lambda1355.5977$ intercombination line, $W_{\rm obs}(3\sigma) \leq 60$ mÅ, which does not help narrow down the range of $N({\rm O}^0)$.

### 3.5. *Al*$^+$

Al II $\lambda$1670.7874 is blended with a C IV doublet at $z_{\rm abs} = 2.5320$ in our 2.9 Å resolution spectrum. However, as can be seen from the bottom panel in Figure 1, subtraction of the C IV absorption does leave a residual feature at the correct wavelength for Al II $\lambda$1670.7874 . Due to the blending, our determination of $N$(Al$^+$) is only accurate to $\pm$0.25 dex.

### 3.6. *Si*$^+$

This is the best-observed element in the present work, with four absorption lines in the echelle spectrum and two in the lower resolution data. The corresponding values of log($\lambda f$) span two orders of magnitude (column 2 of Table 2); consequently our observations constrain both the velocity dispersion of the gas and the column density of Si$^+$ within tight limits. A single absorbing cloud with $b = 10 \pm 1$ km s$^{-1}$ and $N$(Si$^+$)= $(1.4^{+0.5}_{-0.3}) \times 10^{14}$ cm$^{-2}$ gives a satisfactory fit to all six transitions.

However, as discussed by Morton (1991), there are still uncertainties clouding the determination of the relative strengths of different Si II multiplets. In our analysis we have followed the suggestion by Tripp et al. (1995) and adopted the set of $f$-values published by Dufton et al. (1983, 1992). These theoretical calculations appear to be the most reliable at present, because: (i) the value for Si II $\lambda$1808.0126 is in good agreement with the recent measurement by Bergeson & Lawler (1993); and (ii) the values for $\lambda$1304.3702 and $\lambda$1526.7066 are supported by the analysis of interstellar lines recorded at high resolution and signal-to-noise ratio with the *Hubble Space Telescope* (Spitzer & Fitzpatrick 1993). The consequence of using the revised set of $f$-values is to increase the abundance of silicon by about 60% compared to that implied by the $f$-values deduced by Shull, Snow, & York (1981) and adopted in the compilation by Morton (1991); we refer the interested reader to the excellent discussion of this point by Lu et al. (1995).

### 3.7. $S^+$

The strongest line of multiplet 1, S II $\lambda1259.519$, is detected at a $4\sigma$ significance level in our high resolution echelle spectrum (Figure 1 and Table 2). Consequently, the $\pm1\sigma$ limits on $N(\mathrm{S}^+)$ span a factor $\sim 2$ (Table 3). The other members of the multiplet are below our detection limit.

### 3.8. $Fe^+$

Most of the Fe II lines observed are in our intermediate dispersion spectra. However, the one line observed at high resolution, $\lambda1144.9379$, while falling in a noisy part of the spectrum, *is* consistent with both the $b$ value determined from the Si II lines (section 3.6) and the equivalent widths of the other Fe II lines. Consequently, the column density of $\mathrm{Fe}^+$ can be determined to $\pm0.15$ dex.

### 3.9. $Cr^+$, $Ni^+$, and $Zn^+$

None of these species is detected. Nevertheless, our observations place sensitive limits on the corresponding column densities and thereby provide valuable additional information on the pattern of element abundances in the $z_{\mathrm{abs}} = 2.27936$ damped system, which we now discuss.

# 4. A CHEMICALLY YOUNG GALAXY OBSERVED 13 GYR AGO

## 4.1. *Heavy Element Abundances*

The absorption lines recorded are all from dominant ionization stages in H I regions; the corresponding element abundances can therefore be deduced by simply dividing the measured column densities by $N(\mathrm{H}^0)$. However, it is salutary to consider the assumptions underlying this approach.

One source of uncertainty already discussed (section 3.1) is the possibility that only a fraction of the neutral hydrogen may be associated with the heavy elements at $z_{\mathrm{abs}} = 2.27936$, since the Lyman $\alpha$ line encompasses a much wider velocity range than the metal lines (Fig. 2). This effect, if significant, may have led us to underestimate the overall metallicity of the gas, but will not alter the *relative* abundances of different elements.

It is unlikely that we have *underestimated* the total column of neutral gas by neglecting molecular hydrogen, since the molecular fraction in damped Lyman $\alpha$ systems is known to be small (Levshakov et al. 1992). More problematic, at least in principle, is the possibility that some H II gas along the line of sight may be contributing to the first ions, but not to the neutral species, including $\mathrm{H}^0$. This has been a nagging doubt in interstellar abundance studies for twenty years (Steigman, Strittmatter, & Williams 1975). We simply note here that in our data there is no evidence—such as systematic changes in velocity or line profile with ionization potential—for significant amounts of ionized gas near $v = 0$ km s$^{-1}$ in the $z_{\mathrm{abs}} = 2.27936$ system. In general it could be argued that H II contamination is likely to be *less* of a problem in damped Lyman $\alpha$ systems than in Galactic interstellar observations, since the latter normally use OB stars as background sources and therefore of necessity include absorption by gas in the stellar H II region.

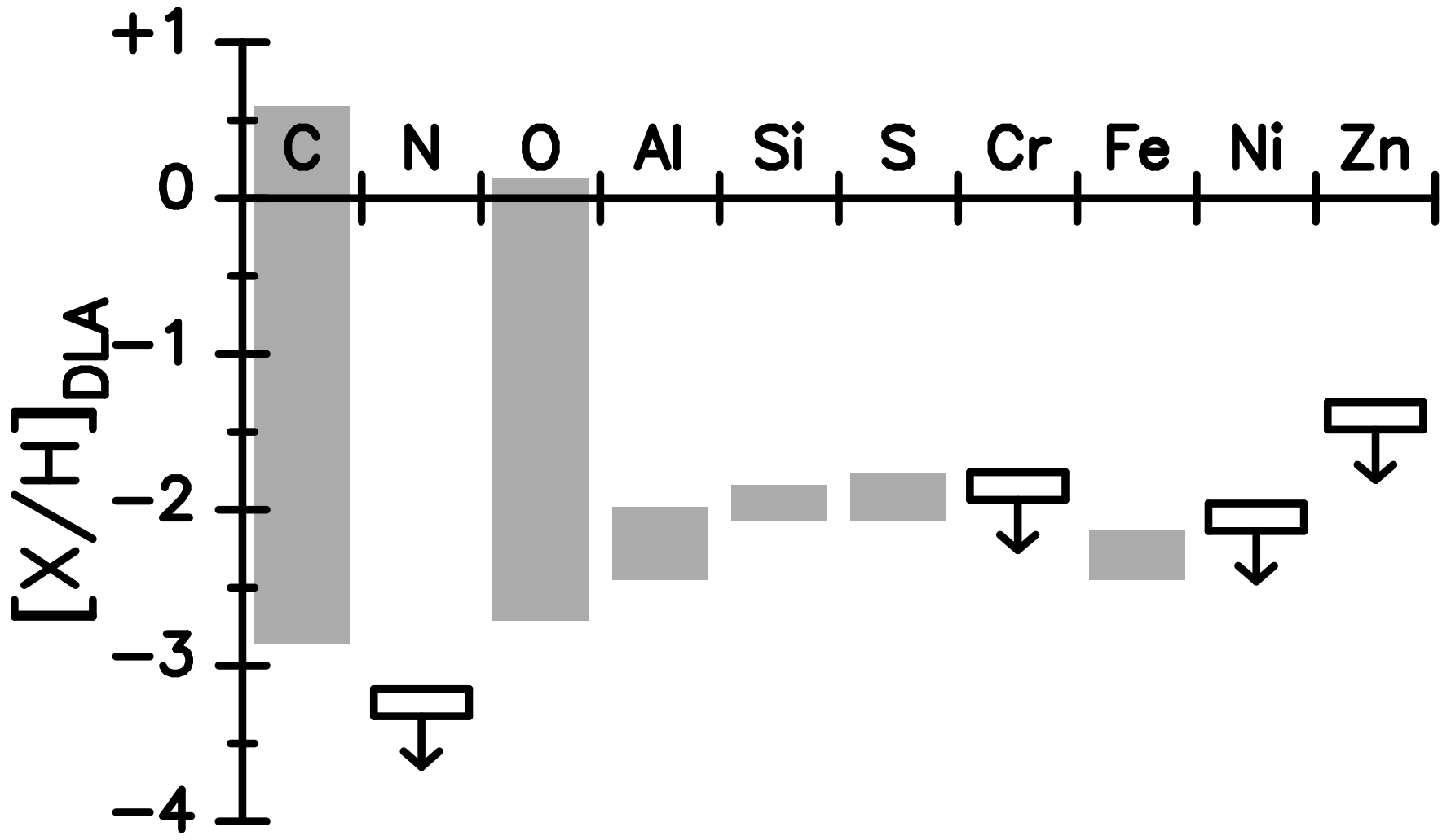

Figure 4: Element abundances in the $z_{\rm abs} = 2.27936$ damped Lyman $\alpha$ system plotted on a logarithmic scale relative to solar values from the compilation by Anders & Grevesse (1989) (thus $[{\rm X/H}]_{DLA} = 0$ corresponds to a solar abundance of element X). Grey vertical bars are for elements with detected absorption lines, the length of the shaded area reflecting the uncertainty in the deduced value of abundance. The abundances of C and O are poorly constrained because the corresponding absorption lines are saturated. Open boxes with downward pointing arrows show upper limits to abundances of elements with absorption lines *below* the detection limits of our data. The upper and lower sides of each box show $3\sigma$ and $2\sigma$ upper limits respectively. The galaxy giving rise to this damped Lyman $\alpha$ system is clearly at an early stage of chemical enrichment, with heavy element abundances only $\sim 1/100$ of solar.

Bearing in mind these potential sources of error, we list in column 5 of Table 3 the element abundances obtained by dividing the values of column density in column 4 by $N({\rm H}^0) = 3.7 \times 10^{20}$ cm$^{-2}$. Comparison with the solar system abundances in column 6, from the compilation by Anders & Grevesse (1989), finally leads to the relative logarithmic abundances $[{\rm X/H}]_{DLA}$ given in column 7 and plotted in Figure 4.

It is clear from these results that the $z_{\rm abs} = 2.27936$ system traces metal-poor gas, with heavy element abundances only $\sim 1/100$ of solar. Evidently, the galaxy responsible for the absorption is still at an early stage of chemical evolution, with perhaps only a few cycles of star formation having contributed to produce the heavy elements we detect. In our own Galaxy metallicities $Z = 1/100 Z_\odot$ are associated with the halo stellar population; by inference we suspect that the damped Lyman $\alpha$ galaxy in front of Q2348−147 had not yet collapsed to form a thin disk at $z \simeq 2$ (if indeed it is the progenitor of a disk galaxy like our own). We note further that, while lower than the 'typical' damped Lyman $\alpha$ metallicity $Z_{DLA} \simeq 1/10 Z_{\odot}$ at $z \simeq 2$ deduced by Pettini et al. (1994), the value we find here for the $z_{\rm abs} = 2.27936$ system in Q2348−147 is neither unusual nor extreme. Several other examples of absorbers with $Z_{DLA} \leq 1/100 Z_{\odot}$ have been reported (Pettini et al. 1995 and references therein).

### 4.2. *Dust Depletion of Refractory Elements*

A second conclusion which is readily drawn from Figure 4 is that dust depletions of refractory elements are significantly reduced compared with those typical of the local interstellar medium. S and Fe are at two extremes in the range of interstellar depletions: while S is undepleted, only $\approx 1\%$ of Fe remains in the gas phase in the cold, diffuse interstellar clouds with values of $N({\rm H}^0)$ comparable to that of the $z_{\rm abs} = 2.27936$ system (Jenkins 1987; Spitzer & Fitzpatrick 1993). In contrast, the *maximum* depletion of Fe relative to S admitted by our measurements is only a factor of $\approx 5$, and *no* depletion is within the $1\sigma$ uncertainties of the abundance determinations. Although less well measured, the abundance of Al—another element which is mostly locked up in grains—is consistent with this picture.

'Typical' damped Lyman $\alpha$ systems with $Z_{DLA} \simeq 1/10 Z_{\odot}$ normally show reduced, *but non-zero*, depletions of elements which readily condense out of the gas, such as Cr and Ni (Pettini et al. 1994 and references therein). It would clearly be of interest to establish if *any* dust has formed in the interstellar medium of galaxies, such as the one considered here, with only 1/100 the metal content of the Milky Way today. From Figure 4 it can be seen that this goal is achievable; a modest increase in S/N should be sufficient to improve the accuracy of the [S/H] and [Fe/H] measurements and the limits on [Cr/H] and [Ni/H] to diagnostically useful levels. Interestingly, Fan & Tytler (1994) and Lu et al. (1995) argued for no depletions of refractory elements in the $z_{\rm abs}$ = 2.8443 DLA system towards the QSO HS 1946+7658, where the overall metallicity is less than 1/100 of solar.

### 4.3. *Overabundance of the $\alpha-$elements*

Metal-poor stars in the halo and disk of our Galaxy show a relative overabundance of O and the even-$Z$ $\alpha$-elements (Mg, Si, S, Ca, and possibly Ti) by about a factor of 3 compared to the Sun (Wheeler et al. 1989; Edvardsson et al. 1993). This well-established observation is a powerful clue to the past history of chemical evolution of the Milky Way. The common interpretation is that the epoch when the overall metallicity of the Galaxy had grown to [Fe/H] $\simeq -1$ corresponds roughly to the evolutionary time scale for the progenitors of Type Ia supernovae which provide an additional source of Fe.

Do we see a similar effect in the interstellar medium of the high-redshift galaxy studied here? The present data are insufficient to answer this question with confidence. Apart from the measurement errors in the individual abundances (particularly O which, as we have seen, is very poorly constrained), we do not have adequate information yet to distinguish the consequences of possible grain depletion from intrinsic departures from solar relative abundances. If we *assume* negligible fractions in the dust, our best estimate is [$\alpha'$/Fe]$\simeq +0.4 \pm 0.2$ where $\alpha'$ is the unweighted average of [Si/H]$_{DLA}$ and [S/H]$_{DLA}$. This value is consistent with that found in metal-poor stars in the Galaxy. The effect of dust depletion would be to reduce this ratio. We thus conclude that there is no evidence in our data for an *enhancement* in the overabundance of the $\alpha-$elements in gas with metallicity $Z_{DLA} \simeq 1/100 Z_{\odot}$.

It is important to extend such measurements to other DLA systems and thereby examine the *distribution* of [$\alpha$/Fe] values at low metallicities. As emphasized by Gilmore

& Wyse (1991), the ratio of $\alpha$ to iron-peak elements may be a discriminant between continuous and sporadic star-formation histories and, if correctly interpreted, has the potential of providing a 'clock' with which to measure the age of the star-burst responsible for the metal enrichment we see at these early epochs.

### 4.4. *The Nitrogen-to-Oxygen Ratio*

We find N to be significantly less abundant that the other elements detected; as can be seen from Figure 4, the effect is so marked that it stands out even with the limited accuracy of the present measurements. For the purpose of the following discussion, it is useful to refer the abundance of N to that of O. From Table 3, we estimate a conservative upper limit [N/O] $\leq -0.44$, based on the lower limit [O/H] $\geq -2.71$ and the $3\sigma$ upper limit from the non-detection of N I lines, [N/H] $\leq -3.15$ . A more sensitive limit can be arrived at by assuming that [O/S] $= 0$; O and S are indeed observed to be in their solar relative proportion in H II regions of all metallicities, down to the lowest values of [O/H] measured (Skillman & Kennicutt 1993; Garnett & Kennicutt 1994). If this is also the case here, we deduce $[\mathrm{N/O}]_{3\sigma} \leq -1.24$, adopting the solar abundances of Anders & Grevesse (1989), or log (N/O) $\leq -2.12$ by number—see Table 3 .

In Figure 5 we compare this value with those found in the H II regions of spiral galaxies—from the extensive compilation by Vila-Costas & Edmunds (1993)—and in dwarf star-forming galaxies, (Pagel et al. 1992; Izotov, Thuan, & Lipovetsky 1994). We have excluded dwarf galaxies with emission line spectra which were judged by the authors to be contaminated by Wolf-Rayet features. The importance of determining the N/O ratio in damped Lyman $\alpha$ galaxies is immediately apparent from Figure 5. Observations such as those presented here probe a region of the (N/O) vs. (O/H) plot which is essentially unexplored because: (i) very few dwarf galaxies with metallicities [O/H] $< -1.5$ (corresponding to 12 + log(O/H) $< 7.43$ in Figure 5) are known, despite recent efforts to track down more examples (e.g. Skillman, Kennicutt, & Hodge 1989), and (ii) the optical [N II] lines are weak and difficult to measure at such low abundances. In contrast, a significant proportion of damped Lyman $\alpha$ systems at $z_{\rm abs} \gtrsim 2$ have $Z_{DLA}$ $\lesssim 1/100$ of solar.

The behaviour of the N/O ratio with increasing metallicity of the gas offers clues to the chemical evolution histories of different galaxies and the stellar populations responsible for producing these two elements. Deciphering these clues from data such as those shown in Figure 5 has been the subject of much debate. Here we limit ourselves to summarizing the basic ideas of a model which has gained favour recently and can explain—at least qualitatively—several aspects of the observations. We refer the interested reader to the reviews by Garnett (1990), Pilyugin (1993), and Vila-Costas & Edmunds (1993) for more detailed treatments.

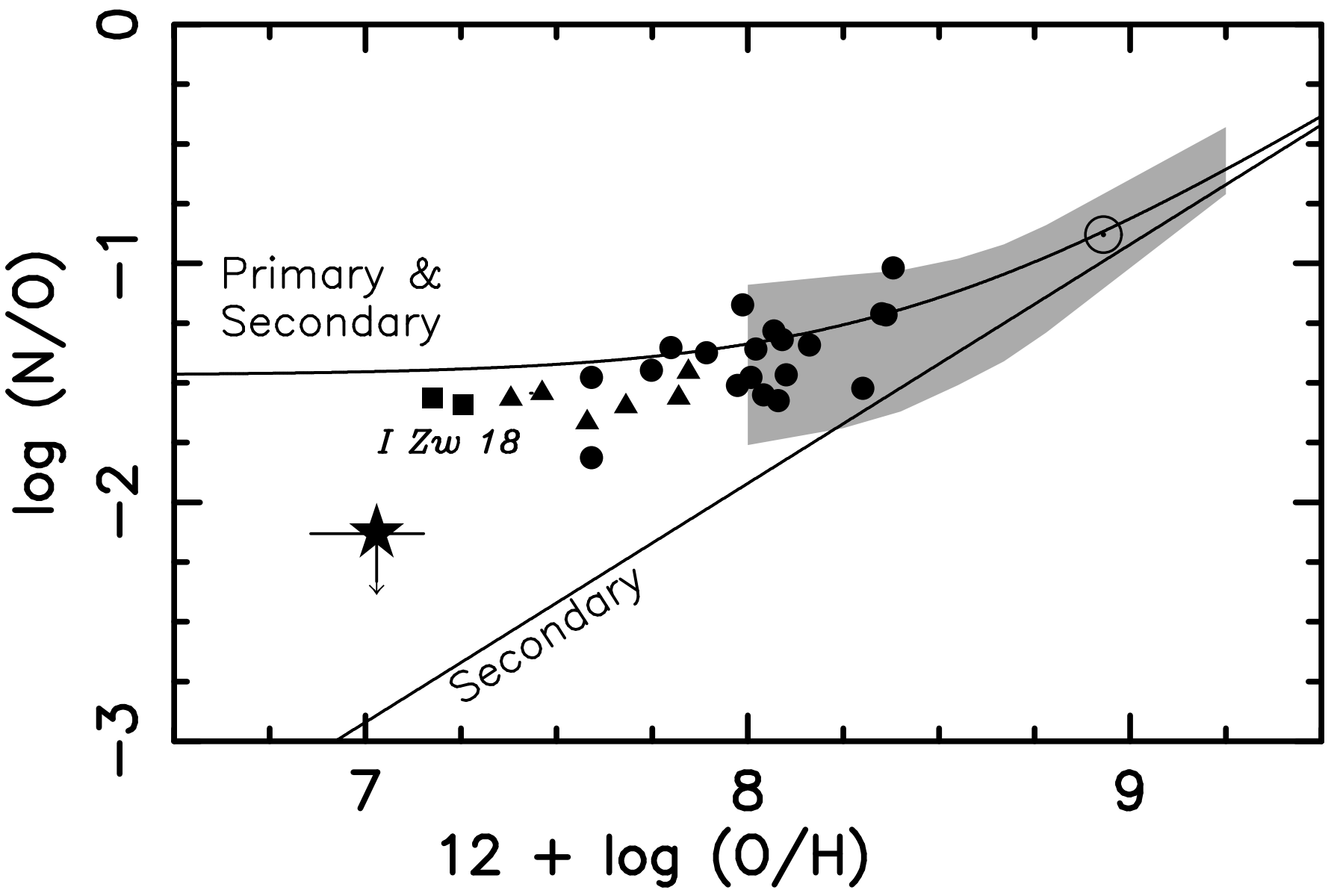

Figure 5: The $3\sigma$ upper limit on the N/O ratio deduced for the $z_{\rm abs} = 2.27936$ damped system assuming O/S to be solar (star symbol) is compared with the values measured in the H II regions of spiral galaxies (shaded area) and dwarf star-forming galaxies (filled symbols). The shaded area includes the central $\approx 3/4$ of the abundance measurements from the extensive compilation by Vila-Costas & Edmunds (1993); the dwarf galaxies selected are those whose spectra Pagel et al. 1992 (circles) and Izotov et al. 1994 (triangles) considered to be free of contamination by Wolf-Rayet features. The two squares labelled I Zw 18 are from the spatially resolved observations of this galaxy by Skillman & Kennicutt (1993). The Sun symbol corresponds to the solar system abundances (Anders & Grevesse 1989). The continuous lines show the best fits to the data determined by Vila-Costas & Edmunds for two possible nucleosynthetic origins of nitrogen, as explained in the text. Our observations of N and O in the damped system extend the study of the relative abundance of these two elements to lower metallicities than reached with the analysis of the emission line spectra of H II regions.

Nitrogen is thought to be synthesized in the CNO processing of oxygen and carbon. If only the O and C originally incorporated into the star when it formed are involved (and a constant mass fraction is processed through a cycle of star formation) then the resulting N is termed 'secondary'. Under these circumstances, (N/O) is expected to increase linearly with (O/H) and such a correlation is indeed observed at high metallicities ([O/H] $\gtrsim -0.6$, or 12 + log (O/H) $\gtrsim 8.3$). An additional source of N must, however, dominate at low metallicities, because in dwarf galaxies the ratio (N/O) exceeds the extrapolation of the linear trend present at high abundances, does not correlate with (O/H), and shows a scatter that is—at least in part—intrinsic, rather than due to observational errors (Pagel 1985). This could be 'primary' N synthesized from the C and O produced by helium burning in the core of the star, and later mixed into a hydrogen burning shell.

The important point for the present discussion is that the major contributors of primary N are thought to be intermediate mass stars ($1-8M_{\odot}$; Renzini & Voli 1981); such stars evolve on much longer time scales than the progenitors of Type II supernovae which are presumably responsible for the bulk of the oxygen production. In this picture, if star-formation proceeds in bursts separated by quiescent periods, (N/O) *increases* with time at a *fixed* (O/H) and all values between pure secondary and primary+secondary are possible.

We now see that the upper limit log (N/O) $\leq -2.12$ which we measure in the $z_{\rm abs} = 2.27936$ system in Q2348$-$147, while lower than the values reported in the most metal-poor H II galaxies, does indeed fall within the expected range in Figure 5. Future observations of other damped Lyman $\alpha$ systems will show if the scatter in the values of (N/O) at a given (O/H) *increases* at the lowest metallicities, as predicted by the model. Furthermore, values of (N/O) close to that for primary production, if found, may allow us to date the onset of star-formation in these primordial galaxies.

## 5. CONCLUSIONS

By bringing together the results of extensive observations of the bright QSO 2348−147, we have measured the abundances of a range of elements in a damped Lyman $\alpha$ system at $z_{\rm abs} = 2.27936$. These data sketch a rough picture of the chemical composition of the interstellar medium in a galaxy which may well be a high-redshift example of galaxies like the Milky Way, observed at a look-back time of $\sim 13$ Gyr. Our main findings are as follows.

1. Heavy element abundances in this system are only $\sim 1/100$ of solar; evidently this galaxy is chemically very young and may have undergone only a few episodes of star formation. While lower than the 'typical' values found at $z \simeq 2$, a metallicity $Z_{DLA} \simeq 1/100 Z_\odot$ is not exceptional, and other such cases are known.

2. The depletions of refractory elements are greatly reduced relative to the local ISM; possibly no dust may have yet formed in this system (or alternatively the grains do not survive in the interstellar conditions prevailing in this galaxy).

3. The ratio of $\alpha$-elements to Fe is in line with the overabundance by a factor of $\sim 3$ seen in Galactic stars of this metallicity.

4. We have measured the N/O ratio at a metallicity *lower* than those of the most metal-poor H II regions known. We find that N/O is less than $\sim 1/15$ of solar, a value which is consistent with current ideas on the relative importance of primary and secondary production of nitrogen at low metallicities. By obtaining data for other high redshift galaxies it will be possible to test these ideas further.

Most importantly in our view, these results demonstrate the potential of damped Lyman $\alpha$ systems in QSOs for complementing, and extending to new regimes, studies of the chemical evolution of galaxies which until now have been based mainly on observations of different stellar populations in our Galaxy and of extragalactic H II regions.

This is a promising area of work which is only now coming within reach and which will surely blossom with the advent of the new generation of large optical telescopes.

We are indebted to David Mar for his competent help with the observations and data reduction, and for enhancing *Xvoigt* to meet the needs of the present analysis. We are grateful to Todd Tripp for communicating his list of recently improved oscillator strengths in advance of publication. It is a pleasure to acknowledge the support of this project by the AAT Time Assignment Committee, and by the AAT and WHT service observing programmes. The interpretation of this work has benefited from useful discussions with Bernard Pagel and Mike Edmunds. K.L. wishes to thank the School of Physics of the University of Sydney for the generous hospitality which made possible his extended visit. R.W.H. gratefully acknowledges financial assistance for this program from the Australian Research Council.